\documentclass[twocolumn]{jpsj3}
\usepackage{cuted,here}
\def\gsim{\mathop {\vtop {\ialign {##\crcr
$\hfil \displaystyle {>}\hfil $\crcr \noalign {\kern1pt \nointerlineskip
 }
$\,\sim$ \crcr \noalign {\kern1pt}}}}\limits}
\def\lsim{\mathop {\vtop {\ialign {##\crcr
$\hfil \displaystyle {<}\hfil $\crcr \noalign {\kern1pt \nointerlineskip
 }
$\,\,\sim$ \crcr \noalign {\kern1pt}}}}\limits}

\usepackage{bm}
\usepackage{color}

\title{Dzyaloshinskii-Moriya Interaction between Multipolar Moments in $5d^1$ Systems}

\author{Masashi {\sc Hosoi}$^1$\thanks{hosoi@hosi.phys.s.u-tokyo.ac.jp}, Tomonari {\sc Mizoguchi}$^{2}$, Taichi {\sc Hinokihara}$^1$, Hiroyasu {\sc Matsuura}$^1$, and Masao {\sc Ogata}$^1$}
\inst
{
$^1$Department of Physics, University of Tokyo, 7-3-1 Hongo, Bunkyo, Tokyo 113-0033, Japan\\
$^2$Departmrnt of Physics, University of Tsukuba, 1-1-1 Tennoudai, Tsukuba, Ibaraki 305-8571, Japan\\
}

\recdate
{
\today
}

\abst
{
We propose a new type of Dzyaloshinskii-Moriya (DM) interactions which act on high-rank multipolar moments such as quadrupolar and
octupolar moments. Here we consider 5$d^1$ systems with broken spatial inversion symmetry, where the interplay of electron correlation,
the spin-orbit coupling, and inversion symmetry breaking plays a crucial role.
Using a numerical diagonalization on a two-site multiorbital Hubbard model, we reveal that anti-symmetric products of multipole
operators have finite expectation values, indicating the existence of DM interactions for multipoles.
We also find that the spin-orbit coupling dependences of DM interactions for multipoles are significantly
different depending on the lattice structure.
Finally, we discuss the numerical results for small and large spin-orbit coupling region by using perturbative analysis.
}
\begin{document}
\sloppy
\maketitle
\section{Introduction}
An interplay of electron correlation and strong spin-orbit coupling (SOC) has attracted much interest due to its novel physical properties.
For electrons in $d$ orbitals, the SOC becomes larger as the principal quantum number increases from $3d$ to $4d$ and $5d$.
In $5d$-based compounds, SOC becomes even comparable with the magnitude of electron correlation.
Thus, they offer an ideal field to investigate the interplay of electron correlation and SOC.

Recently, $5d^5$ systems such as Ir-based magnets have been actively studied.
Examples include Sr$_2$IrO$_4$, which shows an unconventional metal-insulator transition~\cite{Kim},
and Na$_2$IrO$_3$, which is proximity to the Kitaev spin liquid~\cite{Jackeli2009,Singh2010,Rau2014,matsuura1}.
In these materials, Ir$^{4+}$ ions are located at the centre of the octahedral structure, and thus five-fold $5d$ orbitals are split into three-fold
$t_{2g}$ orbitals and two-fold $e_g$ orbitals due to the crystalline electric field~\cite{Sugano1970}.
Then, in the presence of strong SOC, the $t_{2g}$ orbitals with pseudo-orbital degrees of freedom
($L_{\mathrm{eff}}=1$) form upper $J_{\mathrm{eff}}=1/2$ doublet and lower $J_{\mathrm{eff}}=3/2$ quartet,
and only half-filled $J_{\mathrm{eff}}=1/2$ doublet becomes active for $5d^5$ systems~\cite{matsuura1,Witczak2014,Rau2016}.

In contrast to $5d^5$ systems, $J_{\mathrm{eff}}=3/2$ quartet becomes active in $5d^1$ systems.
Remarkably, the exchange interactions between $J_{\mathrm{eff}}=3/2$ states contain not only quadratic operators in $J_{\mathrm{eff}}$,
but also biquadratic and triquadratic operators,
due to its four-fold degrees of freedoms~\cite{Gang}.
These interactions induce many exotic phases
such as the quadrupolar ordered phase in
a double-perovskite material Ba$_2$NaOsO$_6$~\cite{Erickson,Gang}.

It is even more interesting when we consider the effects of spatial inversion symmetry breaking (ISB).
This is because ISB induces the anti-symmetric exchange interaction between magnetic dipole moments, i.e. the Dzyaloshinskii-Moriya (DM) interaction~\cite{Dzyaloshinskii,Moriya}.
For $3d$ systems, the DM interaction has been studied for a long time and the magnitude has been evaluated precisely by first-principles calculations~\cite{Koretune}, and for $5d$ systems,
 the DM interaction between pseudo-spins ($J_{\rm eff}=1/2$) in 5$d^5$ systems also has been
 studied~\cite{Chen2008,Shindou2016,Mizoguchi2016_2,Arakawa2016}.
On the other hand, for $5d^1$ systems, which have the higher-rank multipolar degrees of freedom,
we naturally expect that there exist not only the DM interaction between dipolar moments but also
the analogues of DM interactions between the higher-rank multipoles such as quadrupolar and octupolar moments.

\begin{figure}[b]
\centering
\rotatebox{0}{\includegraphics[width=0.8 \linewidth]{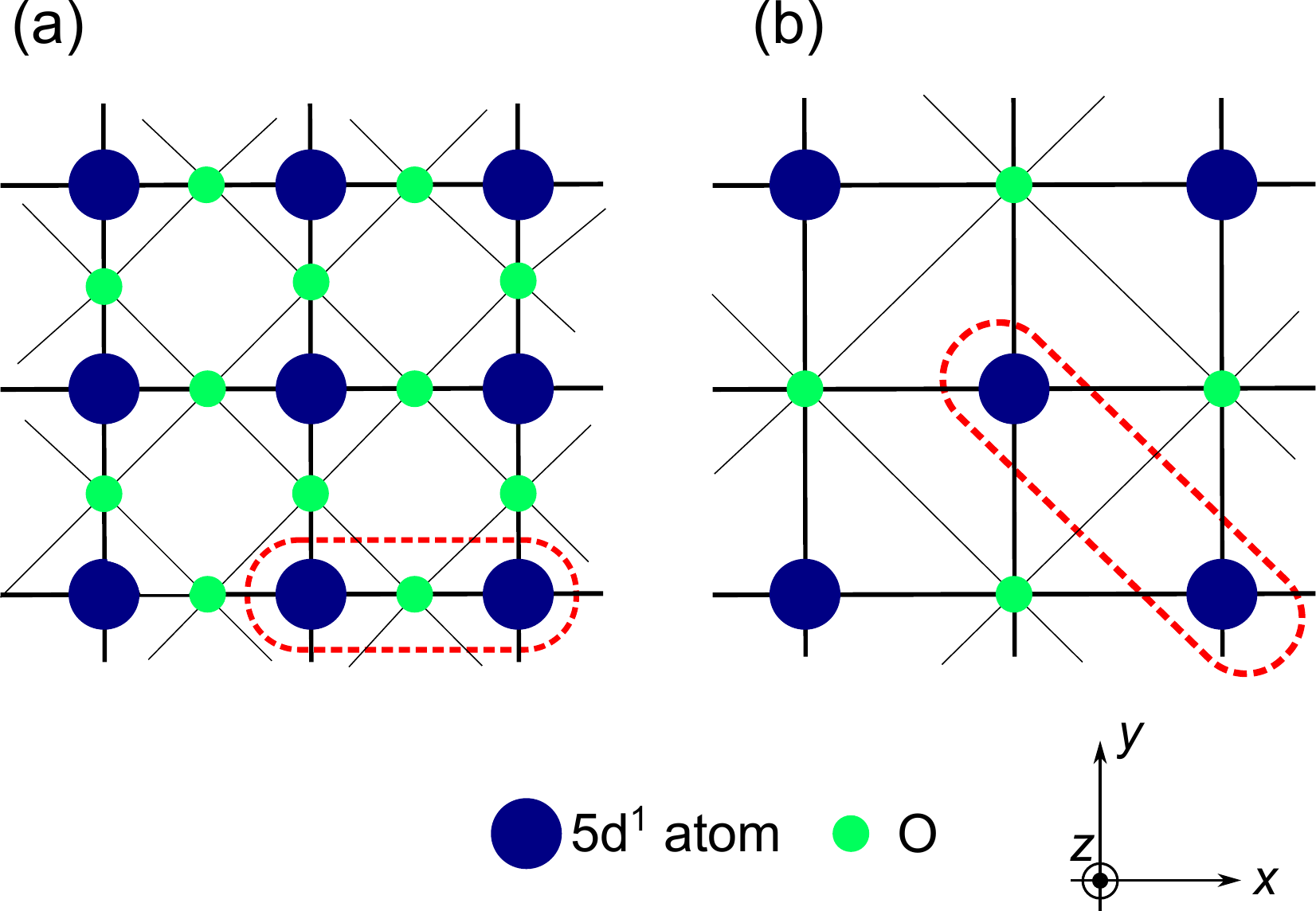}}
\caption{
(Color online) Schematic pictures of (a) corner-sharing configuration and
(b) edge-sharing configuration.
Large navy circles denote $5d^1$ ions and small green circles denote oxygen ions.
}
\label{Fig1}
\end{figure}

In this paper, we clarify the existence of DM interactions for multipolar moments in two-site systems with $t_{2g}$ orbitals and their novel SOC dependences, based on the numerical and analytical approaches.
We also compare two types of perovskite crystals with the corner-sharing and the edge-sharing configurations shown in Figs. \ref{Fig1}(a) and \ref{Fig1}(b) to clarify the structure dependence of the DM interactions between multipolar moments.
Finally, we show that, in the large SOC region, the DM interactions for multipolar moments are finite in the corner-sharing configuration, while these vanish in the edge-sharing configuration
 because of the symmetry requirement of the ground state.

This paper is organized as follows.
In Sect.\ 2, we introduce the model Hamiltonians of two-site systems with the corner-sharing and the edge-sharing configurations, and in Sect.\ 3, we show the numerical diagonalization results on the SOC dependence of DM interactions between multipolar moments in each system.
In Sect.\ 4, we derive effective spin models based on $J_{\mathrm{eff}}=3/2$ state in the limit of small and large SOC, and then compare the difference of DM interactions in two types of configurations.
In Sect.\ 5, we discuss the possibility that multipolar DM interactions vanish even when the inversion symmetry is broken.
Finally, we summarize this paper in Sect. 6.

%---------------------------------------------------------------------------------------------------------%
%---------------------------------------------------------------------------------------------------------%

\section{Model Hamiltonian for 5$d^1$ systems}

%---------------------------------------------------------------------------------------------------------%
\begin{figure}[h]
\centering
\rotatebox{0}{\includegraphics[width=0.8 \linewidth]{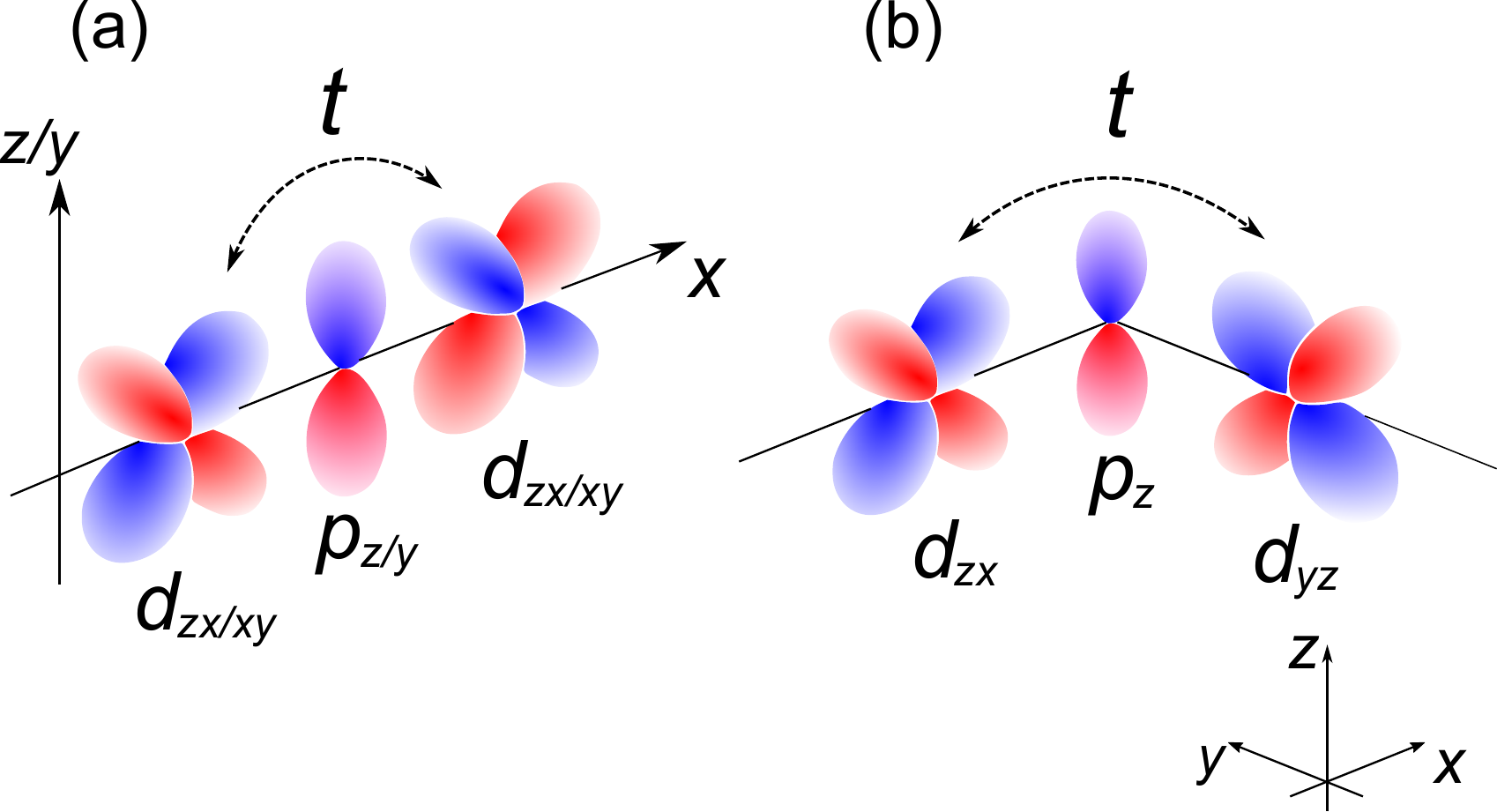}}
\caption{
(Color online) The schematic picture of transfer integral via oxygen's $p$ orbitals in (a) corner-sharing configuration (b) edge-sharing configuration.
$t$ denotes the oxygen-mediated transfer integral between $d$ orbitals. }
\label{Fig2}
\end{figure}
%---------------------------------------------------------------------------------------------------------%

We focus on two sites, which are encircled by red dotted lines in Fig. \ref{Fig1},  in the perovskite structures of the corner-sharing [Fig. \ref{Fig1}(a)] and edge-sharing [Fig. \ref{Fig1}(b)] configurations.
The effective Hamiltonian of this two-site system considering only the $t_{2g}$ orbital is given by
\begin{equation}
  H = H_t + H_{\mathrm{ISB}} + H_{\mathrm{int}} + H_{\mathrm{SO}} , \label{Ham}
\end{equation}
where $H_t$, $H_{\mathrm{ISB}}$, $H_{\mathrm{int}}$, and $H_{\mathrm{SO}}$
represent the Hamiltonians of transfer integrals between $t_{2g}$ orbitals for the inversion-symmetric case, transfer integrals induced by ISB,
on-site Coulomb interactions between $d$ electrons and the SOC in $t_{2g}$ orbitals, respectively.
It is to be noted that two electrons occupy this two-site system since we consider 5$d^1$ configuration.

The Hamiltonian of the transfer integrals, $H_t$, are expressed as
 \begin{equation}
   H_t^{({\mathrm a})} = \sum_{\sigma=\uparrow,\downarrow} t(d_{1,zx,\sigma}^\dagger d_{2,zx,\sigma}^{\phantom{\dagger}} + d_{1,xy,\sigma}^\dagger d_{2,xy,\sigma}^{\phantom{\dagger}} +\mathrm{h.c.}), \tag{2-a}
 \end{equation}
for the corner-sharing configuration, and
 \begin{equation}
   H_t^{({\mathrm b})} = \sum_{\sigma=\uparrow,\downarrow} t(d_{1,yz,\sigma}^\dagger d_{2,zx,\sigma}^{\phantom{\dagger}} + d_{1,zx,\sigma}^\dagger d_{2,yz,\sigma}^{\phantom{\dagger}} +\mathrm{h.c.}), \tag{2-b}
 \end{equation}
for the edge-sharing configuration.
Here, $d_{i,l,\sigma}^\dagger (d_{i,l,\sigma}^{\phantom{\dagger}})$ is a creation (annihilation) operator of the $l$ orbital
($l=yz$, $zx$, and $xy$ orbitals)
with spin $\sigma$ at the $i$-th site, and $t$ is an amplitude of indirect transfer integral between $d$ orbitals derived from the hybridizations between $d$ and $p$ orbitals as shown in Figs. \ref{Fig2}(a) and (b). %, which is the oxygen-mediated transfer integrals.
We ignore the direct transfer integrals between $d$ orbitals ($t_{dd}$) for simplicity.

\begin{figure}[t]
\begin{center}
\rotatebox{0}{\includegraphics[width=\linewidth]{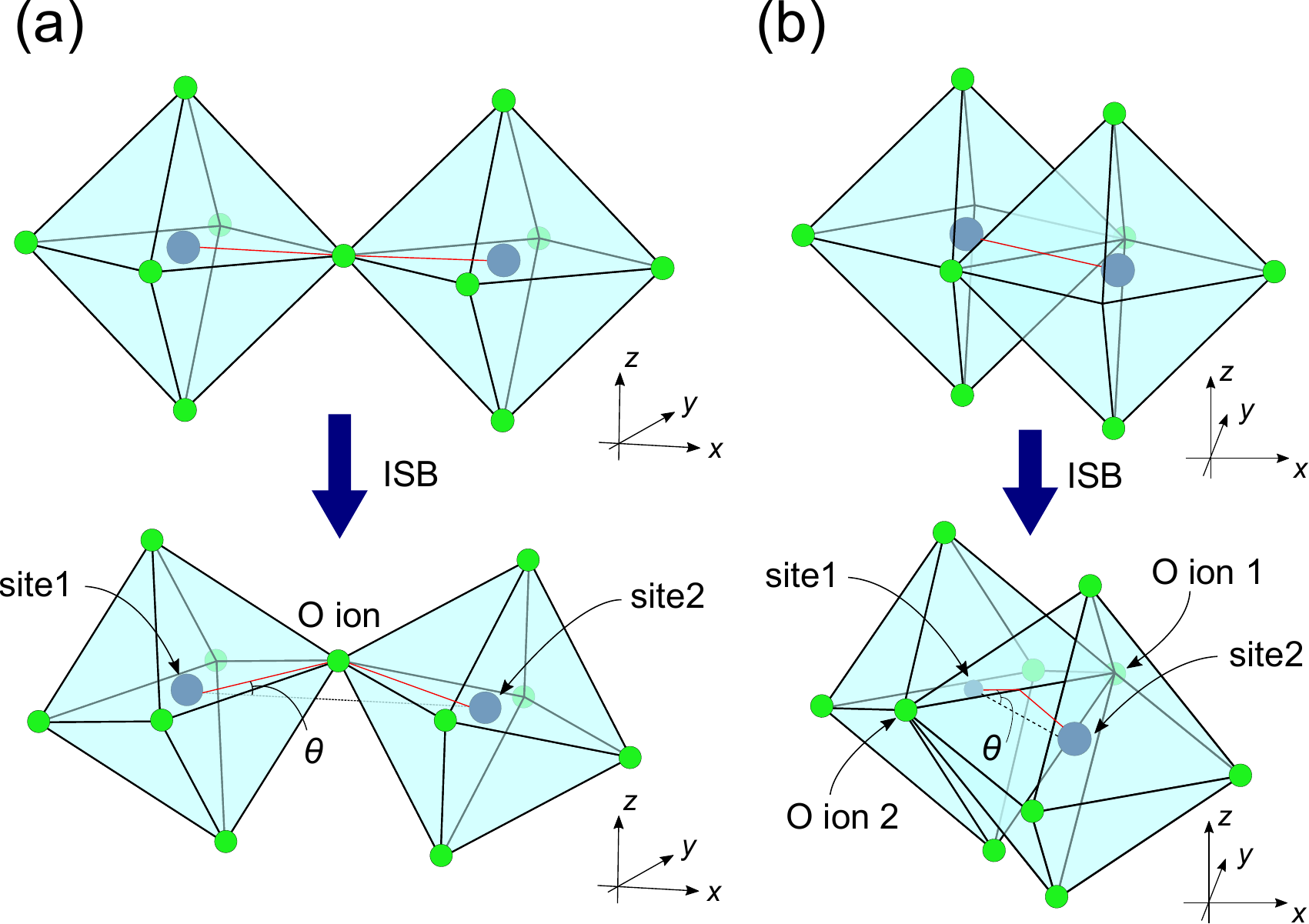}}
\caption{(Color online) The schematic pictures of the configuration of the bond with and without the tilting for (a) corner-sharing configuration and (b) edge-sharing configuration.}
\label{Fig3}
\end{center}
\end{figure}

We study the case where the transfer integrals, $H_{\mathrm{ISB}}$, is induced by the distortion of perovskite structure.
Figure \ref{Fig3}(a) shows the schematic pictures where an oxygen ion (green ball) between the two 5$d^1$ ions (blue balls) shifts slightly in the $z$-direction.
This ISB leads to new hopping processes involving the $d_{yz}$ orbital~\cite{Onoda2002,Yanase2013,Mizoguchi2016_1,Arakawa2016}.
Taking account of the sign of the $d$ orbitals, the Hamiltonian of the transfer integrals induced by ISB becomes
\begin{equation}
  H_{\mathrm{ISB}}^{({\mathrm a})} = \sum_{\sigma=\uparrow,\downarrow} t'(d_{1,yz,\sigma}^\dagger d_{2,xy,\sigma}^{\phantom{\dagger}} - d_{1,xy,\sigma}^\dagger d_{2,yz,\sigma}^{\phantom{\dagger}} + {\rm h.c.}), \tag{3-a}
\end{equation}
for the corner-sharing configuration.
Figure 3(b) shows the case where the two oxygen ions (O ion 1 and 2) shifts slightly in the $z$-direction.
In this edge-sharing case, $H_{\mathrm{ISB}}$ becomes
\begin{align}
  H_{\mathrm{ISB}}^{(\mathrm{b})} = \sum_{\sigma = \uparrow,\downarrow} t''(&d_{1,yz,\sigma}^\dagger d_{2,xy,\sigma}^{\phantom{\dagger}} - d_{1,xy,\sigma}^\dagger d_{2,yz,\sigma}^{\phantom{\dagger}}\notag\\
  &-d_{1,zx,\sigma}^\dagger d_{2,xy,\sigma}^{\phantom{\dagger}} + d_{1,xy,\sigma}^\dagger d_{2,zx,\sigma}^{\phantom{\dagger}} + {\rm h.c.}). \tag{3-b}
\end{align}
The microscopic derivation of $t'$ and $t''$ can be straightforwardly carried out using the Slater-Koster tables~\cite{Mizoguchi2016_1,Slater}.
%-------------------------------------------------------------------------------------%
%---------------------------------------------------------------------------------------------------------%

The Coulomb interaction $H_{{\rm int}}$ and the SOC $H_{{\rm SO}}$, are given by
\begin{align*}
  H_{{\rm int}}
  &= U_d \sum_{i=1,2} \sum_{l} n_{i,l,\uparrow} n_{i,l,\downarrow} \\
  & + \frac{U_{d}^{\prime} - J_{d}}{2} \sum_{\substack{ i=1,2 \\ \sigma}} \sum_{ \substack{ l,m \\ (l \neq m) }} n_{i,l,\sigma} n_{i,m,\sigma} \\
  & + \frac{U_{d}^\prime}{2} \sum_{\substack{i=1,2 \\ \sigma\neq \sigma^\prime}} \sum_{\substack{ l,m \\ (l \neq m) }} n_{i,l,\sigma} n_{i,m,\sigma^{\prime}} \\
  & - \frac{J_{d}}{2} \sum_{i=1,2} \sum_{\substack{ l,m \\ (l \neq m) }} ( d_{i,m,\uparrow}^{\dag} d_{i,m,\downarrow }^{\phantom{\dagger}} d_{i,l,\downarrow}^{\dag} d_{i,l,\uparrow }^{\phantom{\dagger}} \\
  & \hspace{1cm} + d_{i,m,\uparrow}^{\dag} d_{i,m,\downarrow }^{\dagger} d_{i,l,\uparrow}^{\phantom{\dagger}} d_{i,l,\downarrow}^{\phantom{\dagger}} + {\rm h.c.}), \tag{4}
\end{align*}
and
\begin{equation}
H_{{\rm SO}} = \frac{\mathrm{i} \zeta}{2} \sum_{i=1,2} \sum_{\substack{lmn \\ \sigma,\sigma^{\prime}}} \epsilon_{lmn} d_{i,l,\sigma}^{\dagger}
 d_{i,m,\sigma^{\prime}}^{\phantom{\dagger}} \sigma_{\sigma\sigma^{\prime}}^{n}, \tag{5}
\end{equation}
where $n_{i,l,\sigma}$ is the number operator defined as $n_{i,l,\sigma}=d_{i,l,\sigma}^\dagger d_{i,l,\sigma}^{\phantom{\dagger}}$, $\epsilon_{lmn}$ is the Levi-Civita symbol, $\sigma^n$
is the $n$-th component of the Pauli matrices, and $U_d$, $U_d'$, $J_d$, and $\zeta$ are the intra- and inter-Coulomb interactions,
Hund's coupling, and the magnitude of SOC, respectively.
Due to the cubic symmetry, $U_d$, $U_d'$ and $J_d$ satisfy $U_d - U_d' = 2J_d$.

%%%%%%%%%%%%%%%%%%%%%%%%%%%%%%%%%%%%%%%%%%%%%%%%%
\section{Multipolar DM interactions}
\begin{figure}%[b]
\centering
\rotatebox{0}{\includegraphics[width=0.8\linewidth]{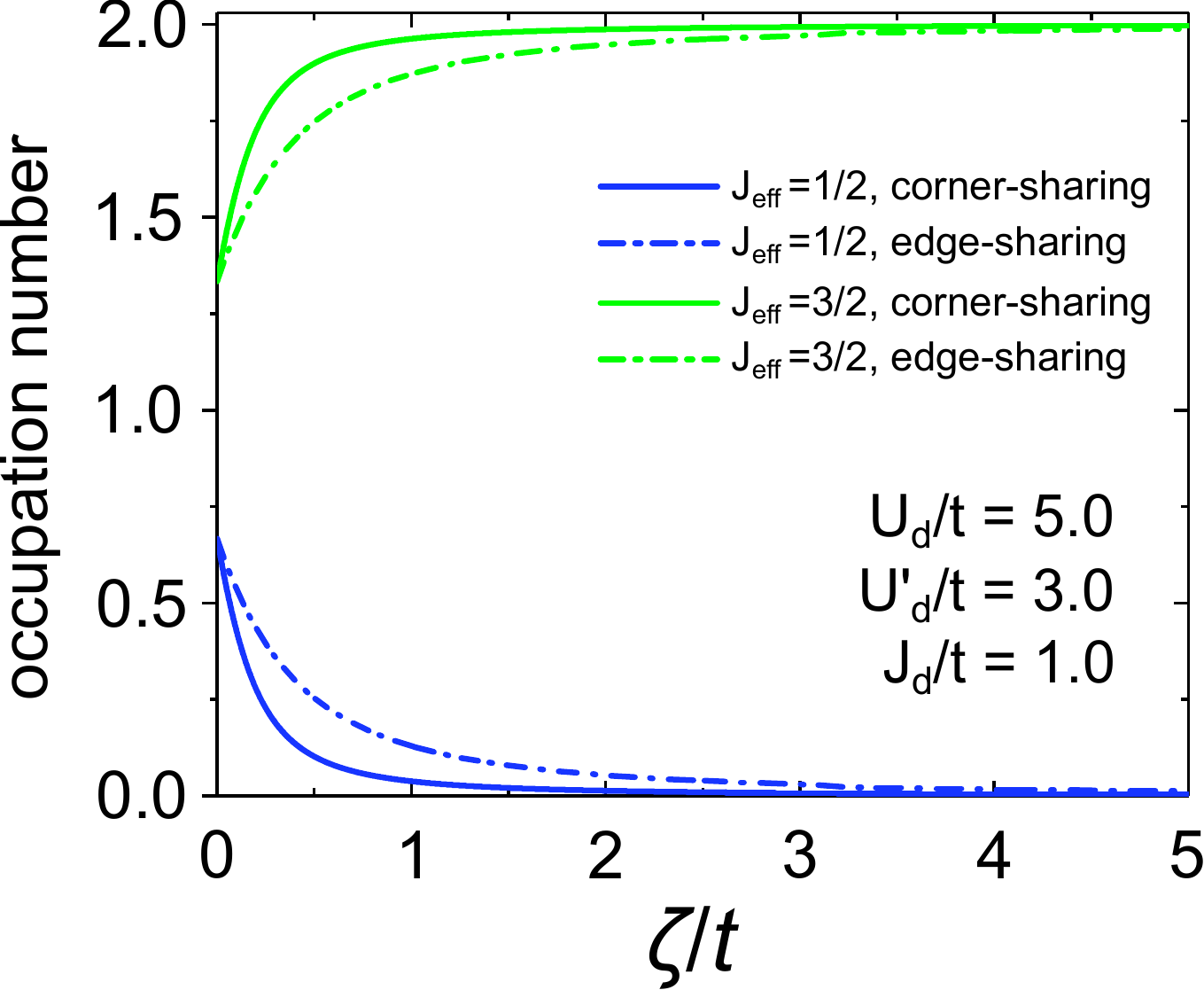}}
\caption{(Color online) SOC dependence of occupation number in $J_{\mathrm{eff}}=3/2$ (green line) and $J_{\mathrm{eff}}=1/2$ states (blue line).
The solid (dot-dashed) line corresponds to the corner-sharing (edge-sharing) configuration.
}
\label{Fig4}
\end{figure}

In order to clarify the multipolar DM interactions between two 5$d^1$ ions, we diagonalize the two-site Hamiltonian Eq.\ (\ref{Ham}) numerically.
First, we calculate the occupation number in $J_{{\mathrm{eff}}} =1/2$ and $3/2$ states.
Here, annihilation operators for $J_{{\mathrm{eff}}} =1/2$ and $J_{{\mathrm{eff}}} =3/2$ states are given by\cite{Sugano1970}
\begin{equation}
  \begin{aligned}
    \phi_{\frac{1}{2},\frac{1}{2}}&=\frac{1}{\sqrt{3}}\left(d_{xy,\uparrow}+d_{yz,\downarrow}-{\mathrm{i}}d_{zx,\downarrow}\right)\\
    \phi_{\frac{1}{2},-\frac{1}{2}}&=\frac{1}{\sqrt{3}}\left(d_{xy,\downarrow}-d_{yz,\uparrow}-{\mathrm{i}}d_{zx,\uparrow}\right),
  \end{aligned}
  \tag{6}
\end{equation}
and
\begin{equation}
  \begin{aligned}
    \phi_{\frac{3}{2},\frac{3}{2}}&=\frac{1}{\sqrt{2}}\left(d_{yz,\uparrow}-{\mathrm{i}}d_{zx,\uparrow}\right)\\
    \phi_{\frac{3}{2},\frac{1}{2}}&=\frac{1}{\sqrt{6}}\left(2d_{xy,\uparrow}-d_{yz,\downarrow}+{\mathrm{i}}d_{zx,\downarrow}\right)\\
    \phi_{\frac{3}{2},-\frac{1}{2}}&=\frac{1}{\sqrt{6}}\left(2d_{xy,\downarrow}+d_{yz,\uparrow}+{\mathrm{i}}d_{zx,\uparrow}\right)\\
    \phi_{\frac{3}{2},-\frac{3}{2}}&=\frac{1}{\sqrt{2}}\left(d_{yz,\downarrow}+{\mathrm{i}}d_{zx,\downarrow}\right).
  \end{aligned}
  \tag{7}
\end{equation}
Here, $\phi_{\alpha,\beta}$ is a annihilation operator of $J_{\mathrm{eff}}=\alpha$ state with $J_z=\beta$.
The occupation number for $J_{{\mathrm{eff}}} =1/2$ and $J_{{\mathrm{eff}}} =3/2$ is defined as
\begin{equation}
  \begin{aligned}
n_{J_{\mathrm{eff}}} &= \sum_{J_z=-J_{\mathrm{eff}}}^{J_{\mathrm{eff}}}\phi_{J_{\mathrm{eff}},J_z}^\dagger \phi_{J_{\mathrm{eff}},J_z}.
\end{aligned}
\tag{8}
\end{equation}
Using the ground state obtained by the numerical diagonalization of Eq. (\ref{Ham}), the occupation number is calculated as $\langle \Phi |n_{J_{\mathrm{eff}}} |  \Phi \rangle$ where $\Phi$ is a wave function of ground sate.
Figure \ref{Fig4} shows the occupation number in $J_{\mathrm{eff}}=3/2$ state (green line) and $J_{\mathrm{eff}}=1/2$ state (blue line) as a function of the magnitude of SOC, $\zeta$, for the case with  $U_d/t=5.0$, $U_d^\prime/t=3.0$, $J_d^\prime/t=1.0$ and $t^{\prime}/t=0$
for the corner-sharing configuration (solid line) and the edge-sharing configuration (dot-dashed line).
The occupation number of $J_{\mathrm{eff}}=3/2$ states ($J_{\mathrm{eff}}=1/2$ states) approaches to 2.0 (0.0) in both configurations for $\zeta/t \gsim 1.0$.
We find that the occupation number is hardly dependent on $t^{\prime}$
at least in the region where $t'\leq0.1t$ is satisfied.
%---------------------------------------------------%
%---------------------------------------------------

As discussed in the introduction, the DM interaction between spins (${S}=1/2$) of $t_{2g}$ orbitals is
expected in the small SOC region ($L$-$S$ region), while the DM interaction between the dipolar moments
in $J_{\mathrm{eff}}=3/2$ state (the dipolar DM interaction) is expected in the large SOC region ($j$-$j$ region).
{To compare the difference between $L$-$S$ and $j$-$j$ pictures, we study the $\zeta$ dependence of
the expectation value of spin and dipolar DM interaction, i.e. $\langle {\bm S}_1\times{\bm S}_2\rangle$ and
$\langle {\bm J}_{1,3/2}\times{\bm J}_{2,3/2}\rangle$.
Here, the operator ${\bm S}_i$ and ${\bm J}_{i,3/2}$ are defined as,
\begin{equation}
  \begin{aligned}
{\bm S}_i&=\frac{1}{2}\sum_{l,\alpha,\beta}d_{i,l,\alpha}^\dagger {\bm \sigma}_{\alpha\beta}d_{i,l,\beta},\\
{\bm J}_{i,3/2}&=\sum_{\alpha,\beta}\phi^\dagger_{i,\frac{3}{2},\alpha}{\bm \tau}_{\alpha\beta}\phi_{i,\frac{3}{2},\beta},
\end{aligned}
\tag{9}
\end{equation}
where the matrix ${\bm \tau}=(\tau^x,\tau^y,\tau^z)$ is
\begin{equation}
  \begin{aligned}
  \tau^x&=
  \begin{pmatrix}
    0 & -\frac{\sqrt{3}}{2} & 0 & 0 \\
    -\frac{\sqrt{3}}{2} & 0 & 1 & 0 \\
    0 & 1 & 0 & \frac{\sqrt{3}}{2} \\
    0 & 0 & \frac{\sqrt{3}}{2} & 0
  \end{pmatrix},\\
  \tau^y&=
  \begin{pmatrix}
    0 & \frac{\sqrt{3}}{2}{\mathrm{i}} & 0 & 0 \\
    -\frac{\sqrt{3}}{2}{\mathrm{i}} & 0 & -{\mathrm{i}} & 0 \\
    0 & {\mathrm{i}} & 0 & -\frac{\sqrt{3}}{2}{\mathrm{i}} \\
    0 & 0 & \frac{\sqrt{3}}{2}{\mathrm{i}} & 0
  \end{pmatrix},\\
  \tau^z&=
  \begin{pmatrix}
    \frac{3}{2} & 0 & 0 & 0 \\
    0 & \frac{1}{2} & 0 & 0 \\
    0 & 0 & -\frac{1}{2} & 0 \\
    0 & 0 & 0 & -\frac{3}{2}
  \end{pmatrix}.
\end{aligned}
\tag{10}
\end{equation}

The expectation values of these DM interaction are calculated using the ground state obtained by numerical diagonalization.
Figure \ref{Fig5} shows the $\zeta$ dependence of the expectation values
$\langle {\bm S}_{1}\times{\bm S}_{2} \rangle_y$ and $\langle {\bm J}_{1,3/2}\times{\bm J}_{2,3/2}\rangle_y$
where we set $t^\prime/t =0.1$ and the other parameters are set as
$U_d/t=5.0$, $U_d^\prime/t=3.0$, $J_d^\prime/t=1.0$.
Note that $x$ and $z$ components vanish for the corner-sharing configuration due to the symmetry requirement. On the
other hand, for the edge sharing configuration, $x$ and $y$ components of this expectation value are exactly the same, and $z$
component vanishes.
At $\zeta/t=0$, $\langle {\bm S}_{1}\times{\bm S}_{2} \rangle_y$ is exactly zero in contrast to $\langle {\bm J}_{1,3/2}\times{\bm J}_{2,3/2}\rangle_y$, and it increases as $\zeta$ increases.
 Besides, we find that its value is much smaller than that between ${\bm J}_{\mathrm{eff}}=3/2$.
Since ${\bm J}_{\mathrm{eff}}$ is given by the combination of spin and orbital angular
momentum, the expectation value of $\langle {\bm L}_1\times{\bm L}_2\rangle_y$ also
contribute to $\langle {\bm J}_{1,3/2}\times{\bm J}_{2,3/2}\rangle_y$.
Thus, the reason why $\langle {\bm J}_{1,3/2}\times{\bm J}_{2,3/2}\rangle_y$ is
finite even at $\zeta/t =0$ of Fig. \ref{Fig5} is due to the orbital component.

%---------------------------------------------------%
\begin{figure}%[h]
\centering
\rotatebox{0}{\includegraphics[width=0.9\linewidth]{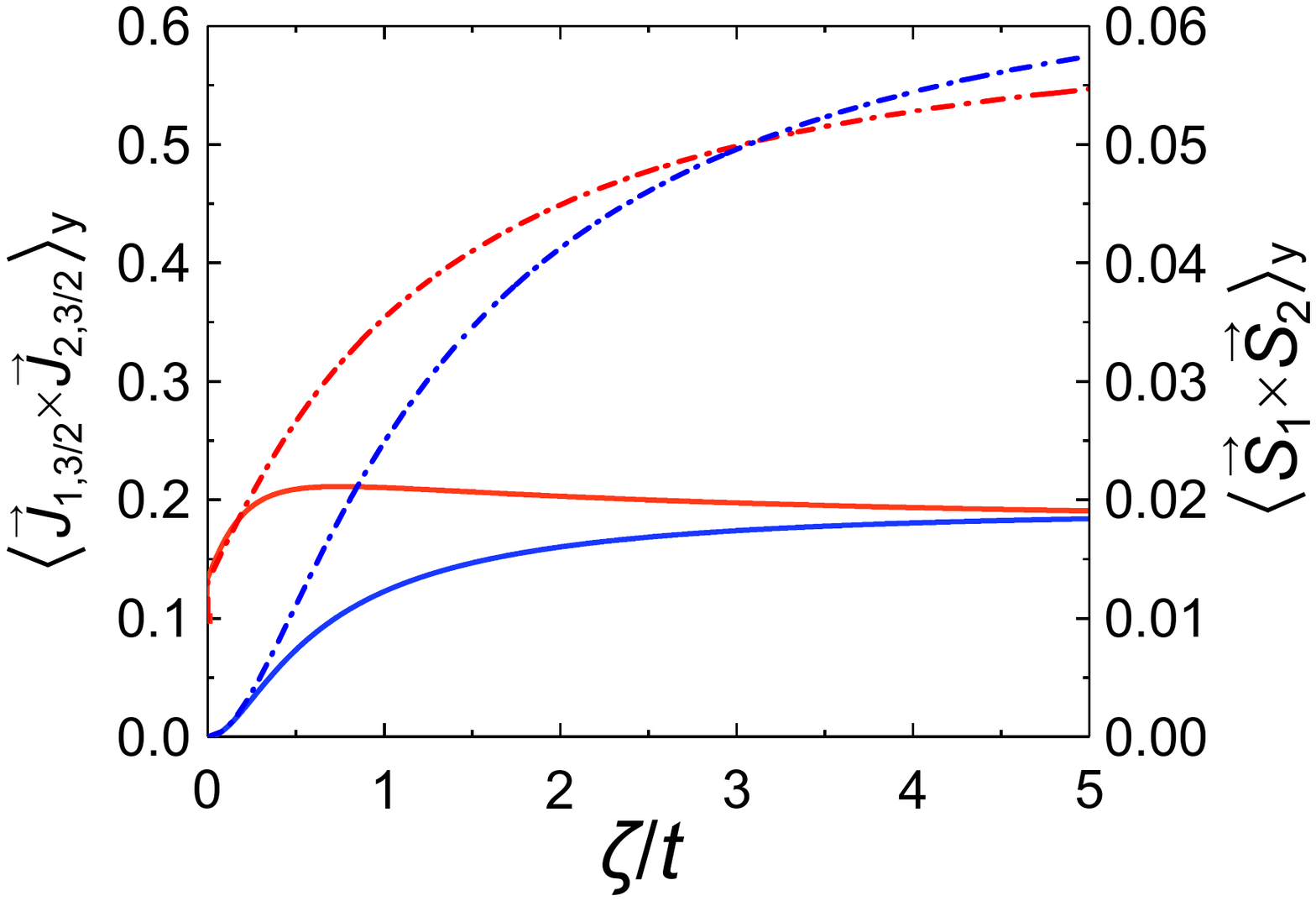}}
\caption{(Color online) The SOC dependence of the expectation value of $y$ component of the DM interaction between ${\bm J}_{3/2}$ states (red), and that between ${\bm S}$ states (blue).
The solid line corresponds to the corner-sharing configuration and the dot-dashed line corresponds to the edge-sharing configuration.
The parameters are set as
$U_d/t=5.0$, $U_d^\prime/t=3.0$, $J_d^\prime/t=1.0$, and $t'/t=0.1$.}
\label{Fig5}
\end{figure}
%---------------------------------------------------%

Figure \ref{Fig6} shows $\zeta$ dependence of the expectation value
$\langle {\bm J}_{1,{3}/{2}}\times{\bm J}_{2,{3}/{2}} \rangle_y$
for $t^{\prime}/t=0.01$ (green line), $0.05$ (blue line) and $0.1$ (red line) in the corner-sharing
configuration (solid line) and edge-sharing configuration (dot-dashed line), respectively.
%---------------------------------------------------%
\begin{figure}%[h]
\centering
\rotatebox{0}{\includegraphics[width=0.9\linewidth]{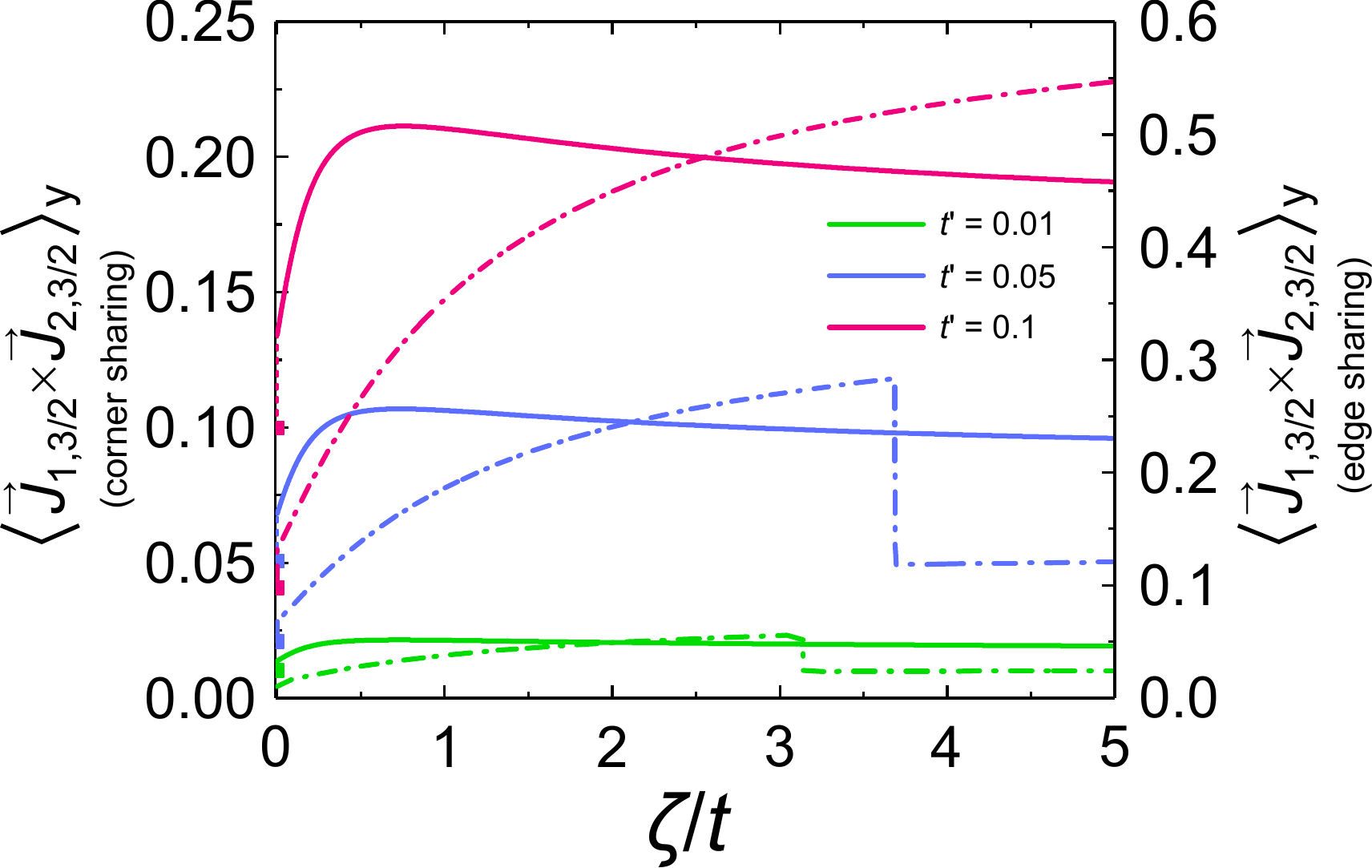}}
\caption{(Color online) The SOC dependence of the expectation value of $y$ component of the DM interaction,
$\langle {\bm J}_{1,{3}/{2}}\times{\bm J}_{2,{3}/{2}} \rangle_y$ for $t^\prime/t =0.01$, $0.05$, and $0.1$.
The solid line corresponds to the corner-sharing configuration (left axis) and the dot-dashed line corresponds to the edge-sharing configuration (right axis).
We use this notation in the following figures.}
\label{Fig6}
\end{figure}
%---------------------------------------------------%
We find that the expectation value $\langle {\bm J}_{1,{3}/{2}}\times{\bm J}_{2,{3}/{2}} \rangle_y$ is finite
in the all of $\zeta$ region of both configurations, indicating the existence of dipolar DM interaction.
In the corner-sharing configuration, this expectation value increases for $\zeta/t \lesssim 0.5$, and then decreases for $\zeta/t \gtrsim 0.5$.
In the edge-sharing configuration, this expectation value increases gradually when $\zeta$ increases, and suddenly drops at certain critical value of $\zeta$.
The origin of this discontinuity will be discussed in {Sect.\ 5}.

Next, let us turn to the higher-rank multipolar degree of freedom.
Possible single-site multipole operators for a cubic $\Gamma_8$ ($J_{\rm eff}=3/2$) state are shown in Table \ref{tab1}~\cite{Gang,Santini,Shiina}.
%%%%%%%%%%%%%%%%%%%%%%%%%%%%%%%%%%%%%%%%%%%%%%%%%%%%%
\begin{table}
  \caption{Multipole moments in a cubic $\Gamma_8$ ($J_{\mathrm{eff}}=3/2$) state. The indices $u$ or $g$ in symmetry representations mean the spatial antisymmetric or symmetric.
   Bracket $[\cdots]$ denotes the symmetrized product of operators in the bracket, e.g. $[J_xJ_yJ_y]=J_xJ_y^2 + J_yJ_xJ_y +J_y^2J_x$.
This table is adapted from Refs. [\citen{Gang,Shiina,Santini}].}
  \label{Table1}
\begin{center}
  \begin{tabular}{lcc} \hline\hline \\
    Multipole & Symmetry & Operator \\ [5pt] \hline \\
    Dipole & $T_{1u}$ & $J_x$, $J_y$, $J_z$ \\ [3pt]
    Quadrupole & $T_{2g}$ & $Q_x$=[$J_yJ_z$]/2  \\ [3pt]
    & & $Q_y$=$[J_zJ_x]/2$  \\ [3pt]
    & & $Q_z$=$[J_xJ_y]/2$  \\ [3pt]
    & $E_g$ & $Q_\alpha$=$J_x^2-J_y^2$  \\ [3pt]
      & & $Q_\beta$=$(2J_z^2-J_x^2-J_y^2)/\sqrt{3}$  \\ [3pt]
   Octupole & $A_{2u}$ & $T_{xyz}=\sqrt{15}/6$ [$J_xJ_yJ_z$]\\ [3pt]
    & $T_{1u}$ & $O_x=J_x^3-{1}/{2}([J_xJ_yJ_y]+[J_xJ_zJ_z])$ \\ [3pt]
    & & $O_y=J_y^3-{1}/{2}([J_yJ_zJ_z]+[J_yJ_xJ_x])$ \\ [3pt]
    & & $O_z=J_z^3-{1}/{2}([J_zJ_xJ_x]+[J_zJ_yJ_y])$ \\ [3pt]
    & $T_{2u}$ & $O'_x={\sqrt{15}}/{6}([J_xJ_yJ_y]-[J_xJ_zJ_z])$ \\ [3pt]
    & &  $O'_y={\sqrt{15}}/{6}([J_yJ_zJ_z]-[J_yJ_xJ_x])$  \\ [3pt]
    & &  $O'_z={\sqrt{15}}/{6}([J_zJ_xJ_x]-[J_zJ_yJ_y])$ \\ [3pt] \hline
  \end{tabular} \label{tab1}
\end{center}
\end{table}
%%%%%%%%%%%%%%%%%%%%%%%%%%%%%%%%%%%%%%%%%%%%%%%%%%%%%
Similarly to the DM interaction between dipoles, the DM interaction between higher-rank multipoles, such as quadrupole and octupole, are expected.
Figure \ref{Fig7} shows $\zeta$ dependence of the anti-symmetric product
of quadrupoles, $\langle {\bm Q}_{1}\times{\bm Q}_{2}\rangle_y$, in the corner-sharing configuration (solid line) and edge-sharing
configuration (dot-dashed line) for $t^{\prime}/t = 0.01$ (green), $0.05$ (blue) and $0.1$ (red), respectively.
Here ${\bm Q}_{i}$ indicates the quadrupolar moment
with $T_{2g}$ symmetry at the $i$-th site. (For details, see Table I.)
%---------------------------------------------------%
\begin{figure}%[h]
\centering
\rotatebox{0}{\includegraphics[width=0.9\linewidth]{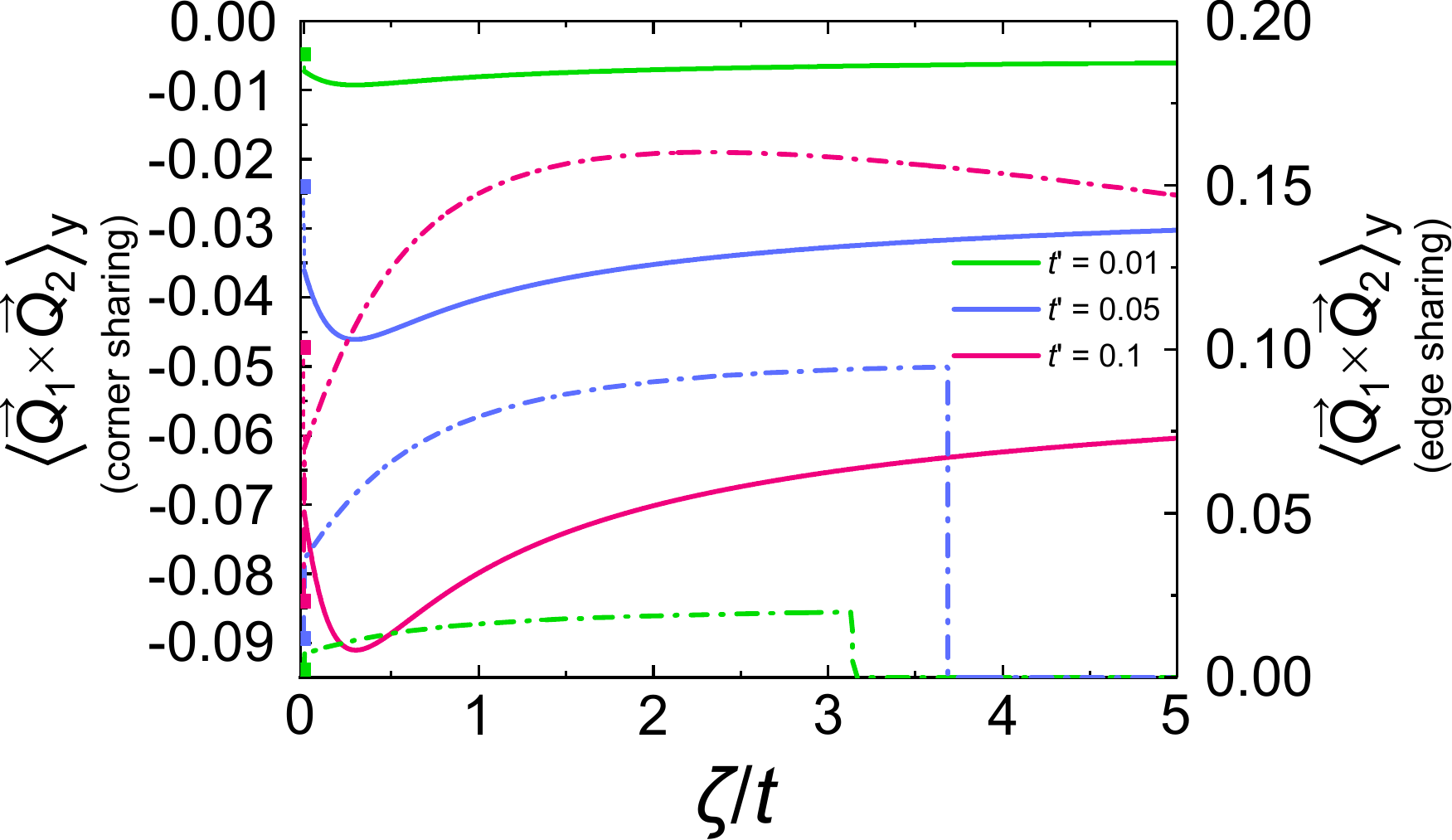}}
\label{graph2}
\caption{(Color online) The SOC dependence of the $y$ component of the interaction ${\bm Q}_{1}\times{\bm Q}_{2}$.}
 \label{Fig7}
\end{figure}
%---------------------------------------------------
It is found that $\langle {\bm Q}_{1}\times{\bm Q}_{2}\rangle_y$ is finite in both configurations, indicating the existence of quadrupolar DM interactions.
Note that it is finite even at $\zeta/t =0$ because the orbital component also takes an important role for higher-rank multipolar DM interactions like in the
case of dipolar DM interaction.
We also find that the expectation value of $\langle {\bm Q}_{1}\times{\bm Q}_{2}\rangle_y$
in the case of the edge-sharing configuration vanishes in the large SOC region above the critical value of $\zeta$.
The origin of this phenomenon will be discussed in Sect.\ 5.

It is to be noted that, in general, the quadrupolar DM interaction can lead to
a lattice distortion through the change of charge distribution.
However, we discuss only the electronic state under the fixed lattice structure in this paper.
The lattice distortion induced by the quadrupolar DM interaction remains as a future problem.

Figure \ref{Fig8} shows $\zeta$ dependence of the octupolar terms, $\langle {\bm O}'_{1}\times{\bm O}'_{2}\rangle_y$
in both configurations with the same parameters as those of Fig. \ref{Fig7},
 where ${\bm O}'_{i}$ indicates the octupolar moment with $T_{2u}$ symmetry at the $i$-th site as shown in Table I.
We find that its expectation value behaves similarly to that of the quadrupole in the region $\zeta/t<1.0$, and vanishes
in the large SOC of the edge-sharing configuration as well as that of Fig. \ref{Fig7}.

%---------------------------------------------------%
\begin{figure}
\centering
\rotatebox{0}{\includegraphics[width=0.9\linewidth]{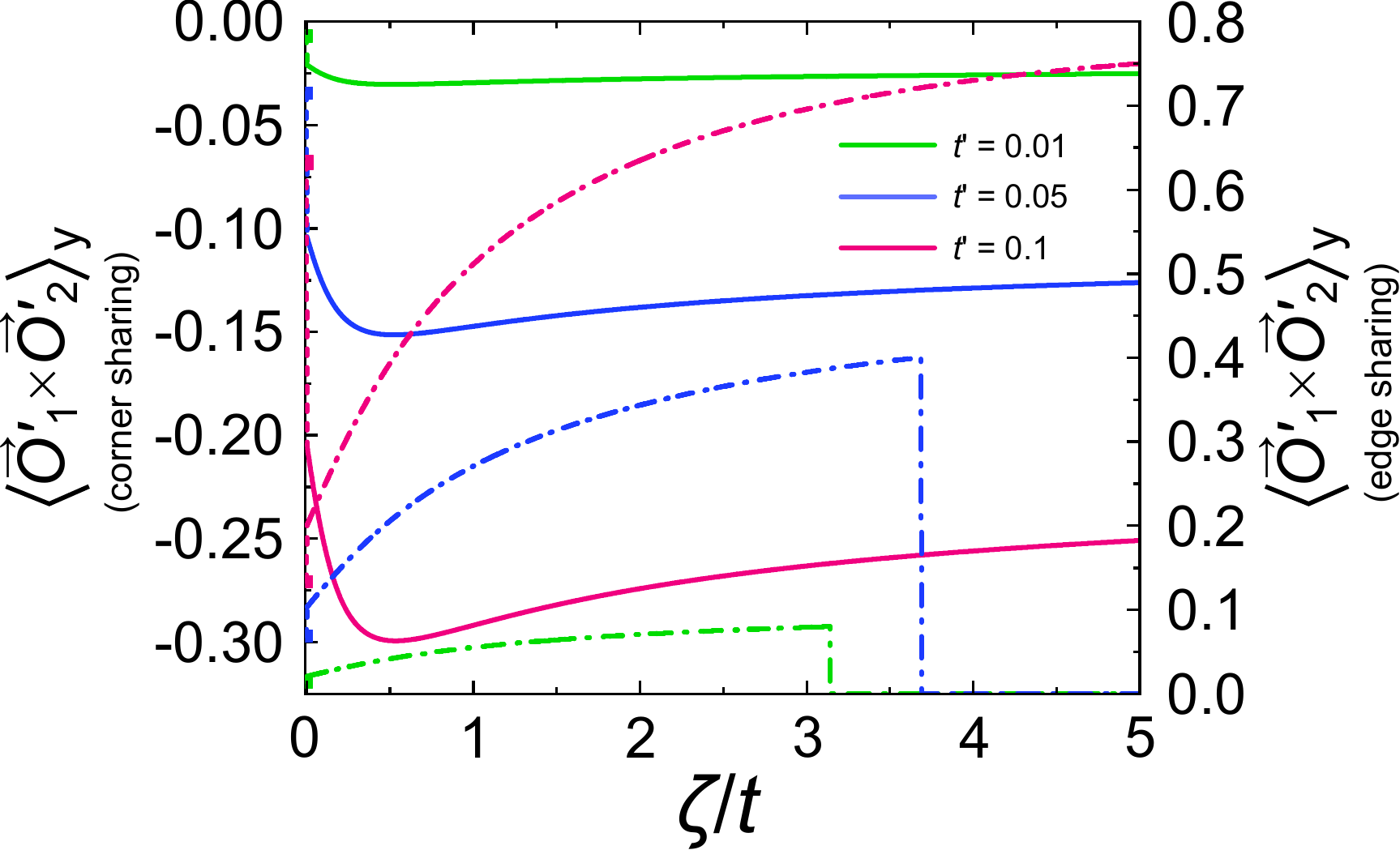}}
\label{graph2}
\caption{(Color online) The SOC dependence of the $y$ component of the interaction ${\bm O}'_1\times{\bm O}'_2$.}
\label{Fig8}
\end{figure}
%--------------------------------------------

Finally, in order to shed light on the feature of multipolar DM interactions,
we show in Fig. \ref{Fig9} the SOC dependence of $\langle {\bm Q}_1\times{\bm Q}_2\rangle_y$ in corner-sharing configuration with several values
of Coulomb interactions. It is found that the DM interaction has a strong dependence of the Hund's coupling $J_d$, namely
 the DM interaction is almost zero at $J_d/t=0$ and it drastically increases as $J_d$ increases.
This is because the wave function of the ground state strongly depends on the value of $J_d$.
Indeed, the ground state is mainly composed of $d_{xy}$ and $d_{yz}$ orbital components at $J_d=0$,
while the contribution of $d_{zx}$ orbital in the ground state increases as $J_d$ increases. It should be noted that
hopping processes including $d_{zx}$ orbital give the main contribution to DM interaction between quadrupoles.
\ \\
\ \\
%---------------------------------------------------%
\begin{figure}[H]
\centering
\rotatebox{0}{\includegraphics[width=0.9\linewidth]{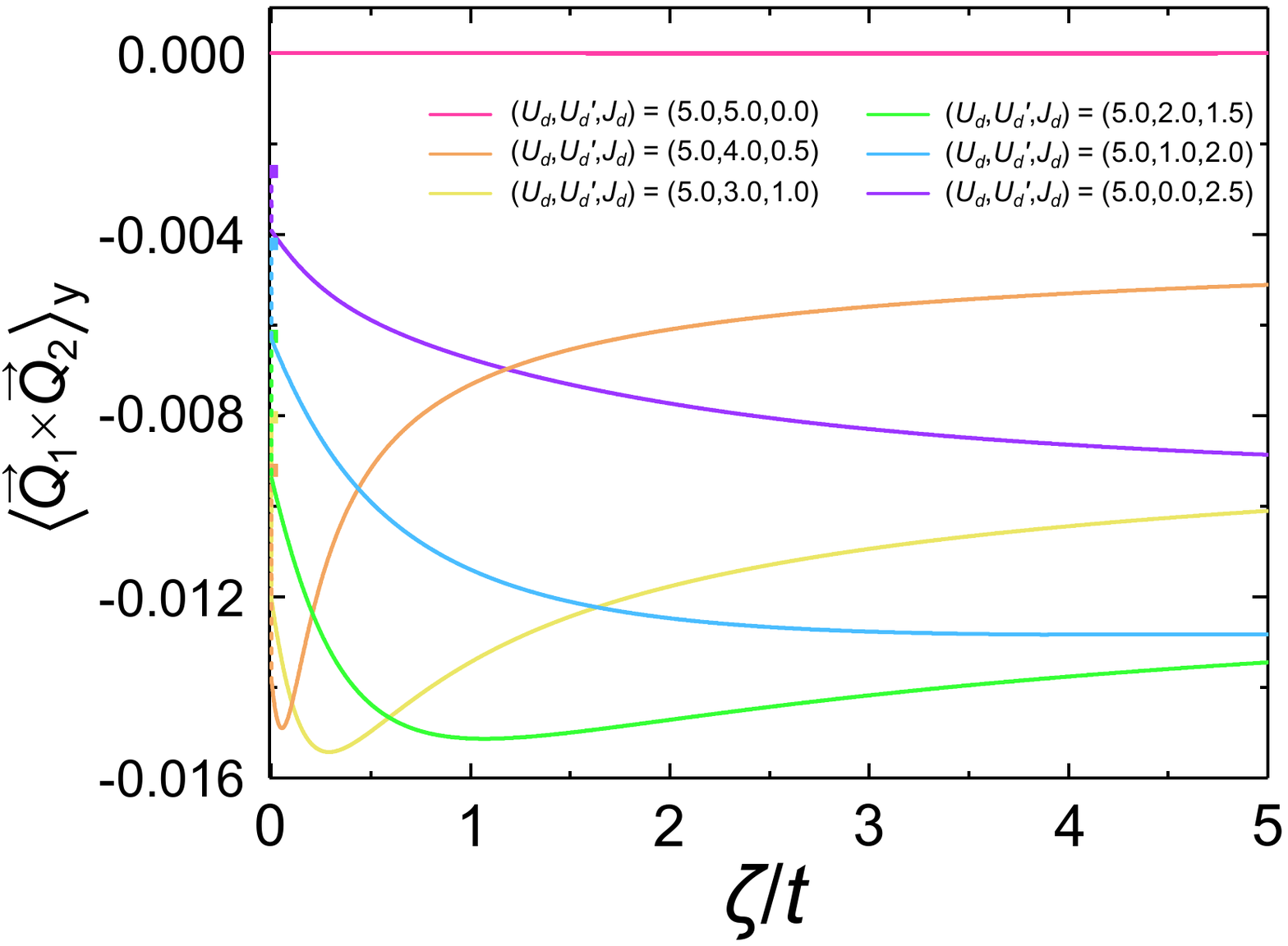}}
\caption{(Color online) The SOC dependence of the $y$ component of the interaction ${\bm Q}_{1}\times{\bm Q}_{2}$ in the corner-sharing configuration for various parameters.
Here, we set $t'/t=0.01$. The expectation values for parameter set ($U_d,U_d',J_d$) = (5.0,5.0,0.0) are small but finite.}
\label{Fig9}
\end{figure}
%---------------------------------------------------%
\section{Analytical results}
In this section, we derive an effective model including multipolar DM interactions
in $J_{\rm eff} = 3/2$ state on the basis of the second order perturbation with respect to the transfer integrals ($H_t + H_{\rm ISB}$),
to clarify the existence and features of these interactions which cannot be captured by numerical diagonalization.

Firstly, we consider the small SOC region.
In the initial state of the second order perturbation process, one electron occupies $J_{\rm eff} = 3/2$ state of each site,
 and then in the intermediate state, two electrons occupy $t_{2g}$ orbitals at the same site.
Since the $L$-$S$ picture is more valid than the $j$-$j$ picture in the small SOC region,
we construct the intermediate states of the effective total angular momentum $J_{\mathrm{eff}}=2,1,0$ with the energies
\begin{align*}
  E_{J_{\mathrm{eff}}=2}&=U_d-3J_d-\zeta, \tag{11-a}\\
  E_{J_{\mathrm{eff}}=1}&=U_d-3J_d+\zeta, \tag{11-b}\\
  E_{J_{\mathrm{eff}}=0}&=U_d-3J_d+2\zeta. \tag{11-c}
\end{align*}
Here, we neglect $J_{\mathrm{eff}}=1/2$ states in the initial state while we treat them in the intermediate states.
We consider that treating $J_{\mathrm{eff}}=1/2$ state in the initial state does not change the qualitative feature
of multipolar DM interactions because the state does not have higher-rank multipolar moment.

By considering the second order perturbation process in the edge sharing configuration, we obtain the effective model for the multipolar DM interactions as
%%%%%%%%%%%%%%%%%%%%%
\begin{strip}
\rule[3em]{\dimexpr(0.5\textwidth-0.5\columnsep)}{0.4pt}
\begin{align*}
{\mathcal{H}}&\sim\frac{2[792(U_d-3J_d)^2+2390(U_d-3J_d)\zeta+1783\zeta^2]}{9(6J_d-2U_d-3\zeta)(3J_d-U_d-2\zeta)(6J_d-2U_d-\zeta)}t't({\bm J}_1\times{\bm J}_2)_y\\
&\phantom{\sim}-\frac{2[32(U_d-3J_d)^2+98(U_d-3J_d)\zeta+71\zeta^2]}{(6J_d-2U_d-3\zeta)(3J_d-U_d-2\zeta)(6J_d-2U_d-\zeta)}t't({\bm Q}_1\times{\bm Q}_2)_y\\
&\phantom{\sim}+\frac{15(-120J_d+40U_d+83\zeta)}{4(3J_d-U_d-2\zeta)(6J_d-2U_d-\zeta)}t't({\bm O}'_1\times{\bm O}'_2)_y+(\text{other interactions}).\tag{12}
\end{align*}
\hfill
\rule[-3em]{\dimexpr(0.5\textwidth-0.5\columnsep)}{0.4pt}
\end{strip}
%%%%%%%%%%%%%%%%%%%%

The effective model for the edge-sharing configuration is also studied in the same way.
Expanding each coupling constant up to the first order with respect to $\zeta$, the coefficients of the first,
second and third terms corresponding to dipolar, quadrupolar and octupolar DM interactions become
\begin{align*}
  -\frac{44t't}{U_d-3J_d}+\frac{389t't}{9(U_d-3J_d)^2}\zeta+{\mathcal{O}}(\zeta^2),\tag{13-a}
\end{align*}
\begin{align*}
  \phantom{-}\frac{16t't}{U_d-3J_d}-\frac{15t't}{(U_d-3J_d)^2}\zeta+{\mathcal{O}}(\zeta^2),\tag{13-b}
\end{align*}
and
\begin{align*}
  \phantom{-}\frac{75t't}{U_d-3J_d}-\frac{255t't}{8(U_d-3J_d)^2}\zeta+{\mathcal{O}}(\zeta^2).\tag{13-c}
\end{align*}
%%%%%%%%%%%%%%%%%%%%%%%%%
From these results, we can see that the dipolar, quadrupolar, and octupolar DM interactions are finite even at $\zeta/t=0$, which is
consistent with those of numerical diagonalization shown in Figs.\ \ref{Fig6}, \ref{Fig7} and \ref{Fig8}.

We also find that the absolute values of these interactions decrease as the SOC increases.
This result seems to be inconsistent with the numerical ones because the absolute values of the expectation values increase as the SOC increases.
However, it should be noted that the expectation values and the coupling constants do not necessarily relate with each other.
As it will be shown in Sect. 5, the lack of direct relationship between them also relates to a mechanism of zero expectation value of multipolar DM interaction in the large
SOC limit.

Next, we derive the effective model in the large SOC region.
Since two electrons occupy only $J_{\mathrm{eff}}=3/2$ states in the initial and intermediate states, it is better to use an effective model restricted only to $J_{\mathrm{eff}}=3/2$ states as follows~\cite{matsuura1}:
\begin{align*}
  H_{\mathrm{int}}&=U_1\sum_{i=1,2}(n_{i,3/2}n_{i,-3/2}+n_{i,1/2}n_{i,-1/2})\\
  &+U_1'\sum_{i=1,2}(n_{i,3/2}n_{i,1/2}+n_{i,3/2}n_{i,-1/2}
  \\&\qquad\qquad\quad +n_{i,-3/2}n_{i,1/2}+n_{i,-3/2}n_{i,-1/2})\\
  &+J_1\sum_{i=1,2}(\phi_{i,3/2}^\dagger\phi_{i,-3/2}^\dagger \phi_{i,1/2}^{\phantom{\dagger}}\phi_{i,-1/2}^{\phantom{\dagger}}+{\mathrm{h.c.}}), \tag{14}
\end{align*}
where $\phi_{i,\alpha}^\dagger(\phi_{i,\alpha}^{\phantom{\dagger}})$ is a creation (annihilation)
operator of $j_z=\alpha$ orbit in $J_{\mathrm{eff}}=3/2$ quartet at the $i$-th site, and $n_{i,\alpha}$ is the number operator defined as $n_{i,\alpha}=\phi_{i,\alpha}^\dagger\phi_{i,\alpha}$.
The parameters $U_1,U_1'$, and $J_1$ are given as
\begin{align*}
  U_1&=F_0+F_2+16F_4, \tag{15-a}\\
  U_1'&=F_0-3F_2-\frac{32}{3}F_4, \tag{15-b}\\
  J_1&=4F_2+\frac{80}{3}F_4, \tag{15-c}
\end{align*}
where $F_0, F_2$, and $F_4$ are Slater-Kondon parameters\cite{Sugano1970}.

The intermediate states are given by the five-fold state ($J_{\mathrm{eff}}=2$) and the singlet state ($J_{\mathrm{eff}}=0$).
Energy levels of these states are obtained as
\begin{align*}
  E_2&\equiv E_{J_{\mathrm{eff}}=2}=F_0-3F_2-\frac{32}{3}F_4, \tag{16-a}\\
  E_0&\equiv E_{J_{\mathrm{eff}}=0}=F_0+5F_2+\frac{128}{3}F_4.\tag{16-b}
\end{align*}
Applying these intermediate states to the the second order perturbation as shown in Fig. \ref{graphhop}, the effective model becomes
\begin{figure*}[t]
\begin{center}
\rotatebox{0}{\includegraphics[width=0.8\linewidth]{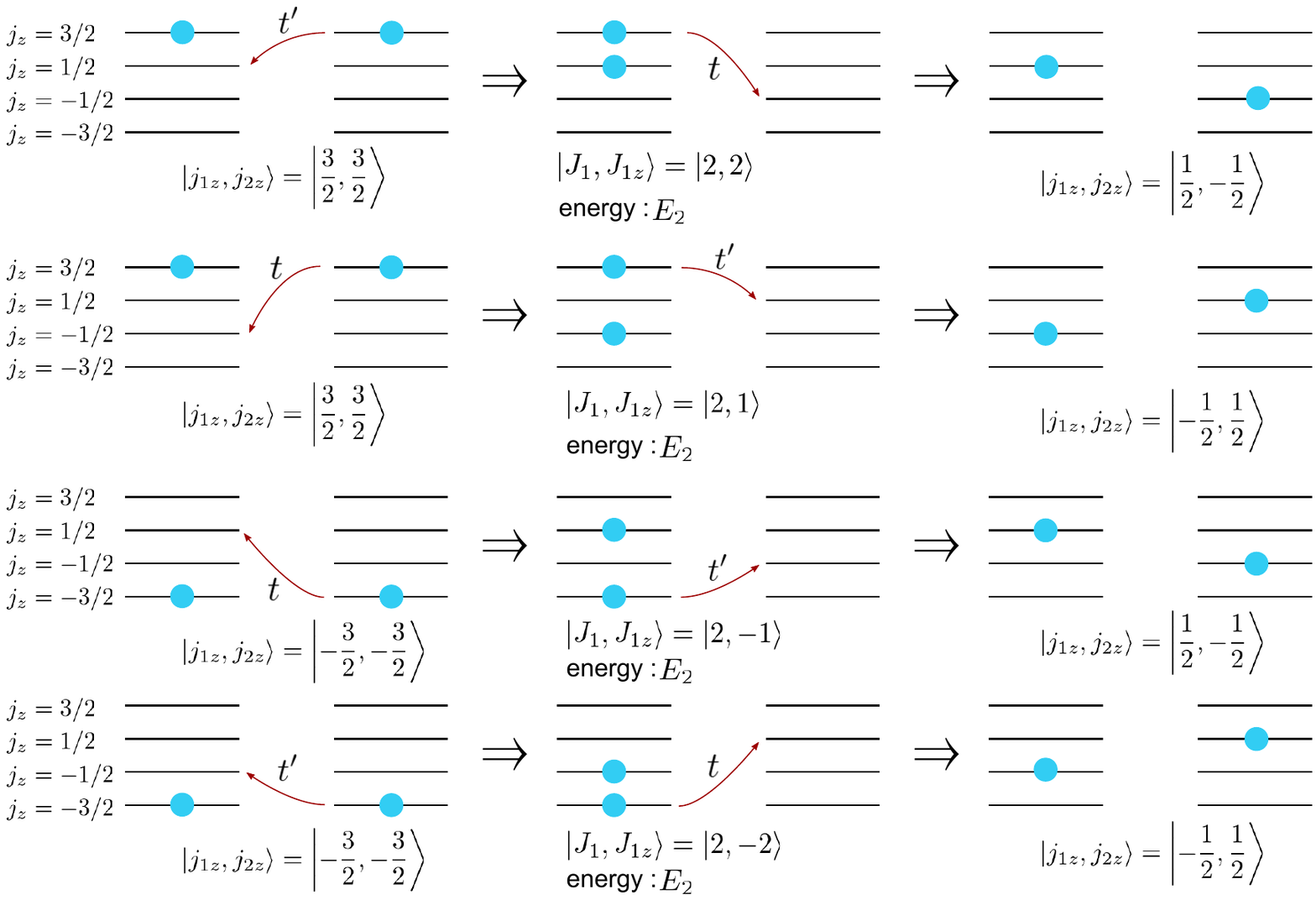}}
\caption{(Color online)The four main perturbation processes for finite expectation value of multipolar DM interactions.
For each processes, the initial and final states are shown with the basis $|j_{1z},j_{2z}\rangle$, and the
intermediate states are shown with the basis $|J_1,J_{1z}\rangle$.}
\label{graphhop}
\end{center}
\end{figure*}
%%%%%%%%%%%%%%%%%%%%%
\begin{strip}
\rule[3em]{\dimexpr(0.5\textwidth-0.5\columnsep)}{0.4pt}
\begin{align*}
{\mathcal{H}}^{\mathrm{(a)}}&\sim -\frac{29E_0+14E_2}{225E_2E_0}t't({\bm J}_1\times{\bm J}_2)_y+\frac{E_0+E_2}{6E_0E_2}t't({\bm Q}_1\times{\bm Q}_2)_y  \\
 &+\frac{E_0-E_2}{135E_0E_2}t't ({\bm O}^\prime_1\times {\bm O}^\prime_2 )_y+\text{(other interactions)}, \tag{17-a}
\end{align*}
for the corner-sharing configuration, and
\begin{align*}
{\mathcal{H}}^{\mathrm{(b)}}&\sim-\frac{2(5E_0+3E_2)}{75E_2E_0}t't[({\bm J}_1\times{\bm J}_2)_x+({\bm J}_1\times{\bm J}_2)_y]+
0[({\bm Q}_1\times{\bm Q}_2)_x+({\bm Q}_1\times{\bm Q}_2)_y]\\
&\qquad\qquad-\frac{8}{135E_2}t't[({\bm O}'_1\times{\bm O}'_2)_x+({\bm O}'_1\times{\bm O}'_2)_y]+\text{(other interactions)},\tag{17-b}
\end{align*}
\hfill
\rule[-3em]{\dimexpr(0.5\textwidth-0.5\columnsep)}{0.4pt}
\end{strip}for the edge-sharing configuration, respectively.
It is found that coupling constants of multipolar DM interactions are finite for both cases in the large SOC region except for the quadrupolar DM interaction for an edge-sharing configuration.
The reason of zero coupling constant is due to the absence of hopping processes corresponding to that interaction.
As shown in Fig. \ref{graphhop}, it is naively expected that the coupling constant of multipolar DM interaction is proportional to $t't/{E_2}$.
In the next section, we compare these analytical results with the numerical results shown in Figs. \ref{Fig6}, \ref{Fig7} and \ref{Fig8}.

%%%%%%%%%%%%%%%%%%%%%%%%%%%%%%%%
\section{Discussion}
Combining the wave function obtained from the numerical calculation and the effective model
derived in the previous section, we discuss the reasons why $\langle{\bm J}_1\times{\bm J}_2\rangle_y$ shows the drastic change,
 and why $\langle{\bm Q}_1\times{\bm Q}_2\rangle_y$ and $\langle{\bm O}'_1\times{\bm O}'_2\rangle_y$ vanish in
  the large SOC region as shown in Figs. \ref{Fig6}, \ref{Fig7} and \ref{Fig8}.

We define $\zeta_c$ as a critical value of SOC where $\langle{\bm Q}_1\times{\bm Q}_2\rangle_y$ and
$\langle{\bm O}'_1\times{\bm O}'_2\rangle_y$ become zero.
For $\zeta < \zeta_c$, the wave function of the ground state is obtained by the numerical calculation as
\begin{align*}
  |\mathrm{GS}\rangle &= ia\left(\left|\frac{3}{2},\frac{3}{2}\right\rangle+i\left|-\frac{3}{2},-\frac{3}{2}\right\rangle\right)\\
  &-be^{i\frac{\pi}{4}}\left(\left|\frac{3}{2},\frac{1}{2}\right\rangle-\left|\frac{1}{2},\frac{3}{2}\right\rangle+\left|-\frac{1}{2},-\frac{3}{2}\right\rangle-\left|-\frac{3}{2},-\frac{1}{2}\right\rangle\right)\\
  &-c\left(\left|\frac{1}{2},\frac{1}{2}\right\rangle-i\left|-\frac{1}{2},-\frac{1}{2}\right\rangle\right)\\
  &-de^{-i\frac{\pi}{4}}\left(\left|\frac{1}{2},-\frac{1}{2}\right\rangle+i\left|-\frac{1}{2},\frac{1}{2}\right\rangle\right). \tag{18}
\end{align*}
Here, we use the basis $|J_{1z},J_{2z}\rangle$. The parameters $a, b, c$, and $d$ complexly depend on the SOC, and increase with lattice distortion. Thus, the expectation values of {$\langle{\bm J}_1\times{\bm J}_2\rangle_y$},
$\langle{\bm Q}_1\times{\bm Q}_2\rangle_y$ and $\langle{\bm O}'_1\times{\bm O}'_2\rangle_y$ change with the SOC.
Note that the components $\left|\frac{3}{2},\frac{3}{2}\right\rangle,\left|-\frac{3}{2},-\frac{3}{2}\right\rangle,\left|\frac{1}{2},-\frac{1}{2}\right\rangle$, and $\left|-\frac{1}{2},\frac{1}{2}\right\rangle$
are essential to obtain the finite values of DM interactions as shown in Fig. \ref{graphhop}.

On the other hand, for $\zeta > \zeta_c$, the wave function of the ground state is
\begin{align*}
  |\mathrm{GS}\rangle &= \tilde{a}e^{i\frac{\pi}{4}}\left(\left|\frac{3}{2},\frac{1}{2}\right\rangle+\left|\frac{1}{2},\frac{3}{2}\right\rangle
  -\left|-\frac{1}{2},-\frac{3}{2}\right\rangle-\left|-\frac{3}{2},-\frac{1}{2}\right\rangle\right)\\
  &-\tilde{b}\left(\left|\frac{3}{2},-\frac{1}{2}\right\rangle+\left|-\frac{1}{2},\frac{3}{2}\right\rangle\right)\\
  &-\tilde{b}i\left(\left|\frac{1}{2},-\frac{3}{2}\right\rangle-\left|-\frac{3}{2},\frac{1}{2}\right\rangle\right). \tag{19}
 \end{align*}
The parameters $\tilde{a}$ and $\tilde{b}$ are independent of the SOC, and $\tilde{b}$ increases with lattice distortion.
By using this state, it is found that $\langle{\bm J}_1\times{\bm J}_2\rangle_y = 48\sqrt{2}\tilde{a}\tilde{b}/25$, and
$\langle{\bm Q}_1\times{\bm Q}_2\rangle_y =\langle{\bm O}'_1\times{\bm O}'_2\rangle_y = 0$.
This disappearance of quadrupolar and octupolar DM interaction is caused by the loss of the components
 $\left|\frac{3}{2},\frac{3}{2}\right\rangle,\left|-\frac{3}{2},-\frac{3}{2}\right\rangle,\left|\frac{1}{2},-\frac{1}{2}\right\rangle$,
and $\left|-\frac{1}{2},\frac{1}{2}\right\rangle$ in the wave function.
For example, for $t'/t = 0.01$, we find $\tilde{a}\simeq0.46, \tilde{b}\simeq0.02$. Thus, we can estimate
$\langle {\bm J}_{1}\times{\bm J}_{2}\rangle_y \simeq 0.025 $, which is consistent with the result of the numerical calculation.
From this calculation, it is found that if the wave function has a symmetric form such as Eq. (18), multipolar DM interactions can be exactly zero even though the lattice is distorted.

{From the perturbative calculation, we have found that the coupling constants of $\langle{\bm J}_1\times{\bm J}_2\rangle_y$ is finite and $\langle{\bm Q}_1\times{\bm Q}_2\rangle_y$ is zero for {the edge}-sharing configuration. This is consistent with the numerical result.
Meanwhile, the coupling constant of $\langle{\bm O}'_1\times{\bm O}'_2\rangle_y$ is finite in the perturbative calculation, which is inconsistent with the numerical result.
In this regard, it should be noted that the finite coupling constant on multipole DM interactions does not necessarily correspond to the finite expectation value of these interactions in the ground state as estimated in Sect. 3.
To compare the numerical result with the perturbative result in detail, the discussion of the ground state based on the effective model of Eq. (17-b) is needed.  }

%%%%%%%%%%%%%%%%%%%%%%%%%%%%%%%
\section{Conclusion}
In conclusion, we have introduced the new type of DM interactions,
namely the quadrupolar and octupolar DM interactions, in $5d^1$ systems
with the structures of edge-sharing configuration and corner-sharing configuration.
Employing the exact diagonalization method for the two-site multi-orbital Hubbard model, we calculated the SOC dependence of
the expectation values of higher-rank multipolar DM interactions, which implies the existence of such interactions.
We also analyzed this model on the basis of the second perturbation theory. As a result, we clarified the reason why
the multipolar DM interactions are finite even in the absence of SOC, and they vanish in the large SOC region in the edge-sharing configuration.

Although the multipolar DM interactions have not been observed yet, there exist some materials in which they can be realized.
For instance, KTaO$_3$ with vacancy of oxygens has an inversion symmetry broken perovskite configuration~\cite{Laguta},
and thus will be a good candidate.
Another possibility is to make a surface/interface of $5d^1$ perovskite material~\cite{Mizoguchi2016_1, Arakawa2016},
on which ISB is artificially introduced.
These novel DM interactions may serve as a source of chiral multipolar orders,
which have not yet been observed experimentally, either.
Therefore, the search for candidate materials will be an intriguing future problem.

Finally, we focus on the Mott insulating phase in this work,
and the comparison with itinerant systems~\cite{Hayami2014,Hitomi2014,Hayami2018} will be an interesting perspective.

{\it Acknowledgement.- }
This work
is supported by the JSPS Core-to-Core Program, A. Advanced
Research Networks. We are also supported by Grants-in-Aid
for Scientific Research from the Japan Society for the Promotion
of Science (Nos. 15K17694, 25220803, 15H02108,
17H02912, and 17H02923 ). M.H. was supported by the Japan Society for the Promotion of Science through Program for Leading Graduate Schools (MERIT).

\end{document}